\documentclass[12pt]{article}
\def\ba{\begin{eqnarray}}
\def\ea{\end{eqnarray}}
\def\vrho{\vec{\rho}}
\def\nn{\nonumber}
\usepackage{graphicx}

\begin{document}
\title{\Huge\bf Heisenberg spins on a cone: an interplay 
between geometry and magnetism} 
\author{{\Large Walter A. Freitas}$^{\,1}$, {\Large Winder A. 
Moura-Melo}\thanks{E-mails: 
winder@fafeid.edu.br,winder@cbpf.br} $^{,2}$ and {\Large A.R.
Pereira}\thanks{E-Mail: 
apereira@ufv.br} $^{,1}$\\ \\ \small 
\it $^1$Departamento de F\'{\i}sica, Universidade Federal 
de Vi\c{c}osa\\ \small\it 36570-000, Vi\c{c}osa, Minas 
Gerais, Brazil.\\ \small \it $^2$ Departamento de 
Ci\^encias B\'asicas, Faculdades Federais Integradas de 
Diamantina\\ \small\it Rua da Gl\'oria 187, 39100-000, 
Diamantina, Minas Gerais, Brazil.}

 \date{}
\maketitle
\begin{abstract}
This work is devoted to the study of how spin texture excitations are 
affected by the presence of a static nonmagnetic impurity whenever they lie 
on a conical support. We realize a number of novelties as 
compared to the flat plane case. Indeed, by virtue of the conical shape, 
the interaction potential between a soliton and an impurity appears to 
become stronger as long as the cone is tighten. As a consequence, a new 
kind of solitonic excitation shows up exhibiting lower energy than in the 
absence of such impurity. In addition, we conclude that such an energy 
is also dependent upon conical aperture, getting lower values as the 
latter is decreased. We also discuss how an external magnetic field 
(Zeeman coupling) affects static solitonic textures, providing instability 
to their structure.\end{abstract}


\section{Introduction and Motivation}

Two-dimensional (2D) Heisenberg-like spins models has attracted a great 
deal of efforts in the last decades. Actually, they have been applied to 
study a number of magnetic materials displaying several 
properties\cite{ref1}. For instance, the continuum version of the isotropic 
2D Heisenberg lattice (so-called nonlinear $\sigma$ model- NL$\sigma$M) is 
useful for investigating properties of quasi-planar isotropic magnetic 
samples in the long-wavelenth and zero-temperature limits. In this case, 
the nonlinearity of the continuum theory supports static excitations with
associated finite energy, 
the so-called Belavin-Polyakov solitons\cite{BP}. Such pseudo-particles, 
like the kinks of sine-Gordon or the 't Hooft-Polyakov monopoles of 
Georgi-Glashow model have their stability guaranted by topological features 
of the model rather than the equations of motion (for a review, see, for 
example Ref.\cite{Ryder}). In fact, the presence of topologically stable 
nonlinear excitations in magnetic systems may render the latter 
ones interesting new phenomena. As a example, we quote the role played by 
vortex pair dissociation which induces a topological phase transition in 
the sample, at the temperature $T_{\rm BKT}$, even though no long range 
order is established\cite{BKT}.\\

On the other hand, even the purest samples contain impurities (and 
defects, imperfections, etc) whose presence may give rise to important new 
properties. This is the case, for instance, of 
artificially dopped semiconductor materials. In magnetic materials, 
impurities may be considered for improving (magnetic impurity) or for 
vanishing (nonmagnetic, spinless) local magnetic interaction at the 
positions where they were placed. However, recent reported results have 
found that the antiferromagnetic correlations around a spin vacancy are not 
destroyed, rather they appear to be increased
\cite{Julienetal,Subetal,Zaspeletal}. Such a result has leads to
the appearance of a new type of soliton whose energy is 
lower than its counterpart in the absence of the spinless 
impurity\cite{Subetal,Zaspeletal,PereiraPires}. This `more fundamental
soliton' has been theoretically
\cite{Zaspeletal,PereiraPires,Moletalprb2002,nosprb,Moletalprb2003} studied
and observed in experiments\cite{Subetal}.\\

In the work of Ref.\cite{PereiraPires}, the NL$\sigma$M supplemented by a 
static nonmagnetic impurity potential was applied to study such a new 
soliton. In spite of the investigation be carried out in the low frequency 
limit, its results were shown to be in excellent agreement with numerical 
and discrete lattice ones \cite{Subetal,Zaspeletal}. Actually, as presented in
Ref.\cite{PereiraPires}, 
an attractive potential takes place between the soliton and the spin 
vacancy as long as they are sufficiently close. In addition, such a 
potential is expected to give rise to oscillating solitons with definite 
frequencies around the static impurity \cite{nosprb} and could be taken as 
a trapping mechanism for solitonic excitations\cite{Afraniopla}. \\

The discussion presented above is mainly concerned to magnetic systems 
defined on a flat plane, i.e., curvatureless manifold. Nevertheless, recent 
works have dealt with Heisenberg-like spins on curved background, e.g., 
cylinders, cones, spheres, thorus. In these cases, a number of new 
phenomena have been described, like the geometrical frustration on spin 
textures induced by curvature and/or by non-trivial topological aspects of the 
space manifold, say, angular deficit in cones, area deficit in 
planes with a disk cut out, and so forth (see, for example, 
Refs.\cite{Saxenaphysa,curvatura1,curvatura2,AfranioJMMM}). Actually, the
study of 
such systems may be of considerable importance for practical applications, 
for example, in connection to {\em soft condensed matter} 
materials\cite{Saxenaphysa} (deformable vesicles, membranes, etc), and 
also to artificially nanostructured curved objects (nanocones, 
nanocylinders, etc) in high storage data devices\cite{Ross1,nanocones}.\\

Here, we shall consider the Heisenberg spin system lying on a conical 
background in the presence of a static spinless impurity. As we shall see, 
by virtue of an interplay between geometry (cone aperture angle, $\alpha$) 
and magnetism, soliton-like excitations experience a collective effect of 
impurity and geometry so that, the potential between both objects appears 
to become stronger as the cone is tighten. As a consequence, we also 
realize that new solitonic excitations appear exhibiting lower energy as 
$\alpha$ is decreased. We discuss on possible consequences of these results 
whenever compared to the flat plane case. Furthermore, we analyse the 
effect of an external magnetic field on static solitons and show that such 
a coupling becomes these excitations unstable.\\

\section{The model and the influence of spin vacancy on solitonic excitations
on the cone} 

We shall start by considering the continuum limit of the 2D isotropic
Heisenberg spins model supplemented by the presence of a static spin vacancy
potential like below:

\ba
H=\frac{J}{2}\,\int\,d\rho^2 \left[(\partial_t \vec{n})^2-(\nabla 
\vec{n})^2\right]\,V(\vec{\rho})\,,\label{Hini}
\ea
with $J>0$ representing the antiferromagnetic coupling (in the static
limit, $\partial_t\vec{n}=0$, and with $J<0$, the
model above could describe a ferromagnetic system), while the integral is
evaluated over a
conical surface with coordinates $\vec{\rho}=(\rho, \tau)$ related to the
usual cylindrical ones, $(r,\phi)$, by:
\ba
\rho=\frac{r^\beta}{\beta}\,,\qquad \tau=\beta\varphi\,.\nn
\ea
In addition, $\vec{n}=[(\sin\theta \cos\Phi;
\sin\theta \sin\Phi; \cos\theta)]$ is the N\'eel spin vector state, with
independent variables $\theta(\vec{\rho},t)$ and $\Phi(\vec{\rho},t)$.
Furthermore, following the work of Ref.\cite{PereiraPires}, we may write the
non-magnetic impurity, $V(\vec{\rho})=V_I(\vec{\rho})$, as:

\ba
V_I(\vec{\rho})=\left\{\begin{array}{l}0\quad 
{\rm if} |\vec{\rho}-\vec{\rho}_I|<A\\  1\quad {\rm if} 
|\vec{\rho}-\vec{\rho}_I|\ge A \end{array} \right. 
\,,\label{Vvac}
\ea
with $A=a^\beta/\beta$, $a$ is the lattice spacing parameter.\\

Then the dynamical equations which follows from Hamiltonian (\ref{Hini}) read
like (hereafter, $m=\cos(\theta)$):

\ba
& & \frac{\partial\,m}{\partial\,t} =V_I(\vec{\rho})\left[\nabla^2\Phi 
-\frac{2m}{(1-m^2)}(\nabla 
m)\cdot(\nabla\Phi)\right]-\nabla\Phi\cdot\nabla 
V_I(\vec{\rho})\,,\label{eqdphi} \\ & & 
\frac{\partial\,\Phi}{\partial\,t}= V_I(\vec{\rho})\left[\nabla^2 m
+\frac{m}{(1-m^2)}(\nabla 
m)^2 +m(1-m^2)(\nabla\Phi)^2\right] -\nabla
m\cdot\nabla V_I(\vec{\rho})\,. 
\label{eqdm}
\ea

A straighforward calculation also gives\footnote{For circular conical
coordinates, $(\rho,\tau)$, we have that: $$\nabla
f(\rho,\tau)=(\beta\rho)^{(1-1/\beta)}\left[\hat{e}_\rho\,
\frac{\partial}{\partial\rho} f + \frac{\hat{e}_\tau}{\rho}\,
\frac{\partial}{\partial\tau} f\right]\quad{\mbox{\rm and}}\quad \nabla^2
h(\rho,\tau)= (\beta\rho)^{2(1-1/\beta)}
\left[\frac{\partial^2}{\partial\rho^2}\, h +\frac{1}{\rho}
\frac{\partial}{\partial\rho}+\frac{1}{\rho^2}
\frac{\partial^2}{\partial\tau^2}\right].$$}:

\ba
\nabla V_I(\vec{\rho})= (\beta\rho)^{1-1/\beta}\,
\frac{a^\beta}{\beta}\left[\hat{e}_\rho 
\cos(\gamma-|\tau-\tau_I|)+\hat{e}_\tau\sin(\gamma-|\tau-
\tau_I|)\right]\delta(\vrho-\vrho_I-\vec{A})\label{gradVI}\,.
\ea
 
As may be easily checked, whenever we take the static limit in eqs. above in
the absence of the spin vacancy (thus, $V(\vec{\rho})=1$ everywhere in
Hamiltonian (\ref{Hini}), etc), such eqs. are solved by static spin textures
lying on the
cone with the profiles (see Ref.\cite{AfranioJMMM}):

\ba
m(\vec{\rho})=m_0(\rho)=\frac{\rho^2 -\rho_0^2}{\rho^2+\rho_0^2}\,,\qquad 
\Phi(\vec{\rho})=\Phi_0(\tau)=\tau\label{BPsolcone}\,,
\ea
which represent solitonic excitations with radius
$R=(\beta\rho_0)^{1/\beta}$ and associated energy $E_0=4\pi\,J\beta$. Thus, in
the $\beta\to1$ limit (usual planar case), the results above recover those
describing the so-called Belavin-Polyakov solitons,
presenting radius $R$ and energy $E_{BP}=4\pi\,J$ (see Ref.\cite{BP}). Then,
solutions (\ref{BPsolcone}) are the conical counterparts of the
Belavin-Polyakov solitons.\\

As a first result, notice that such objects present lower energy
whenever lying on a cone. Thus, if we consider a planar surface with one or
more imperfections in conical shape (cusps, etc) then it is expected that
solitonic excitations will tend to nucleate around the apices of such cusps,
once for presenting such a configuration a smaller amount of energy is
demanded. This could be
viewed as a geometrical pinning of solitons.\\

\begin{figure}[ht] 
\centering \hskip -2cm
\includegraphics[width=11cm,height=8cm]{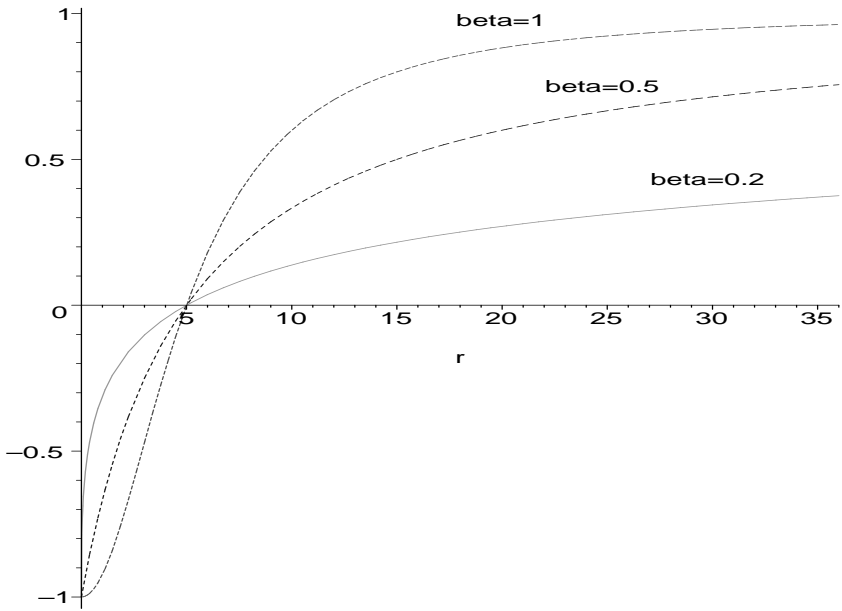}
\vskip -0.0cm \caption{{\protect\small Shows solitonic profile $m=m_0$ as
function of distance, $r$, centered at the apex cone
($\vec{r}=\vec{0}=\vec{\rho}$). Note how such a behavior changes as $\beta$
decreases. Here, we have taken $R=5a$.}} \label{fig1} 
\end{figure}

In addition, we should notice that the magnetization variable, $m$, behaves in
a peculiar way on the cone whenever compared to the flat plane case.
Figure \ref{fig1} shows how $m=m_0$ varies with distance, $r$. Notice that
as $\beta$ decreases (cone is tighten) spins located from the apex to $r=R$
(or, equivalently $\rho=\rho_0$), present stronger variations than those
outside soliton radius, $r>R$. Physically, such an asymmetry in the variations
comes about by the area deficit (brought about by the angular deficit): as
$\beta$ gets
smaller the area on the cone, say, from the apex to $\rho_0$, also decreases,
lefting minor space to spins turn from $m_0=-1$ to $m_0=0$. In the limit of
very small $\beta$, magnetization should abruptly pass from $m_0=-1$ to
$m_0=0$, while the change from $m_0=0$ to $m_0=1$  would be performed in a so
smoothly way.\\

As a first step forward, let us see how spin vacancy affects static solitons.
For that, let us consider that the interaction of spins with such a static
impurity is much less than spin-spin coupling. Then, we shall take:
$m(\vec{\rho})=m_0(\rho)+m_1(\vec{\rho})$ and $\Phi(\vec{\rho})=\Phi_0(\tau)+
\Phi_1(\vec{\rho})$ where $(m_0, \Phi_0)$ are given in (\ref{BPsolcone}), while
$(m_1, \Phi_1)$ are small corrections to the former ones, provided by $V_I$. In
this case, we may linearize eqs. (\ref{eqdm}) with respect to $m_1$ and
$\Phi_1$ for $\rho\ge\,a^\beta/\beta$, obtaining:

\ba
& & \nabla^2\Phi_1
-2\frac{m_0\vec{\nabla}\,m_0}{(1-m_0^2)}\cdot\vec{\nabla}\Phi_1=
-\frac{a^\beta}{\rho_I}\,\sin(\gamma)\delta(\vec{\rho}
-\vec{\rho_I})\,, \label{eqfi1} \\
& & \nabla^2\,m_1 +2\left[
\frac{(\nabla\,m_0)^2+m_0\vec{\nabla}m_0\cdot \vec{\nabla}}{(1-m_0^2)}
\right]m1
+2m_0(1-m_0^2) \vec{\nabla}\Phi_0\cdot\vec{\nabla}\Phi_1=\nn\\ 
& &
\hskip
7cm -\frac{4\rho_0^2\rho_I\,a^\beta}{(\rho_I^2+\rho_0^2)^2}\,\cos(\gamma)\,
\delta(\rho-\rho_I)\,.
\label{eqm1}
\ea
Now, considering that the spin vacancy is located at the soliton center, say,
$\vec{\rho_I}=(0,0)$ (conical apex), then $\sin(\gamma)=0$,
yielding $\Phi_1={\rm constant}$ for eq.
(\ref{eqfi1}), while (\ref{eqm1}) becomes:

\ba
\nabla^2m_1+2\frac{(\nabla\,m_0)^2+m_0\vec{\nabla}\,m_0}{(1-m_0^2)}
\cdot\vec{\nabla} \, m_1 = 0\,,\nn
\ea
or still, by recalling that the 1st derivative term disappears if we change
to $m_0\equiv {u}$, we get:

\ba
\frac{d^2\,m_1}{du^2}+\frac{2\,m_1}{(1-u^2)}=0\label{eqm1u}\,,
\ea
whose solutions are the trivial one, $m_1=0$ and:
$$
m_1(m_0)= c_1(1-m_0^2) +c_2\left[ m_0+\frac12
(1-m_0^2)\,ln\left(\frac{1+m_0}{1-m_0}\right) \right]\,,
$$
with $c_1$ and $c_2$ being constants to be determined by the solitonic
characteristics. Indeed, by demanding that as $\rho\to\infty$, $m=m_0+m_1\to1$
we must impose $c_2=0$. Then, the complete solutions in the presence of the
spin vacancy are the following:

\ba
& & m_P=m_0\,, \quad \Phi_P=\Phi_0\label{Psoliton}\,,\\
& & m_I=m_0+c_1(1-m_0^2) \,,\quad \Phi_I=\Phi_0 \label{Isoliton}\,.
\ea
Actually, we should notice that the presence of the nonmagnetic impurity
support both of the two solitonic excitations above, similarly to what happens
even in the usual plane case, Ref.\cite{PereiraPires}. Their energy read as
below:

\ba
E_j=\frac{J}{2}\int_0^{2\pi\beta} \,d\tau
\int_{a^\beta/\beta}^\infty \left[ \frac{[\vec{\nabla}(m_0+m_j)]^2}{(1-m_0^2)}
+(1-m_0^2)(\vec{\nabla} \Phi_0)^2 \right]\rho\,d\rho\,,\label{energysoliton}
\ea
where $j$ stands for soliton type, say, $j=I,P$, while the lower cut-off in the
$\rho$-integral, $a^\beta/\beta$, takes into account the effect of
$V_I(\rho)$-potential. Evaluating these integrals
over the cone, we finally obtain (recall $E_0=4\pi\,J\beta$):

\ba
& &
E_I=E_0\left[\frac{R^{4\beta}+(Ra)^{2\beta}(1-c_1)}{(R^{2\beta}+a^{2\beta})^2}
\right]
\label{energysolitonI}\\
& & 
E_P=E_0\left[\frac{R^{2\beta}}{(R^{2\beta}+a^{2\beta})}\right]\,.
\label{energysolitonP}
\ea
Now, minimizing $E_I$ with respect to soliton size, $R(a)$, we see that
$0<c_1\le 1/2$, analogously to its counterpart found in Ref.
\cite{PereiraPires}.\\

\begin{figure}[ht] 
\centering \hskip -2cm
\includegraphics[width=13cm,height=9cm]{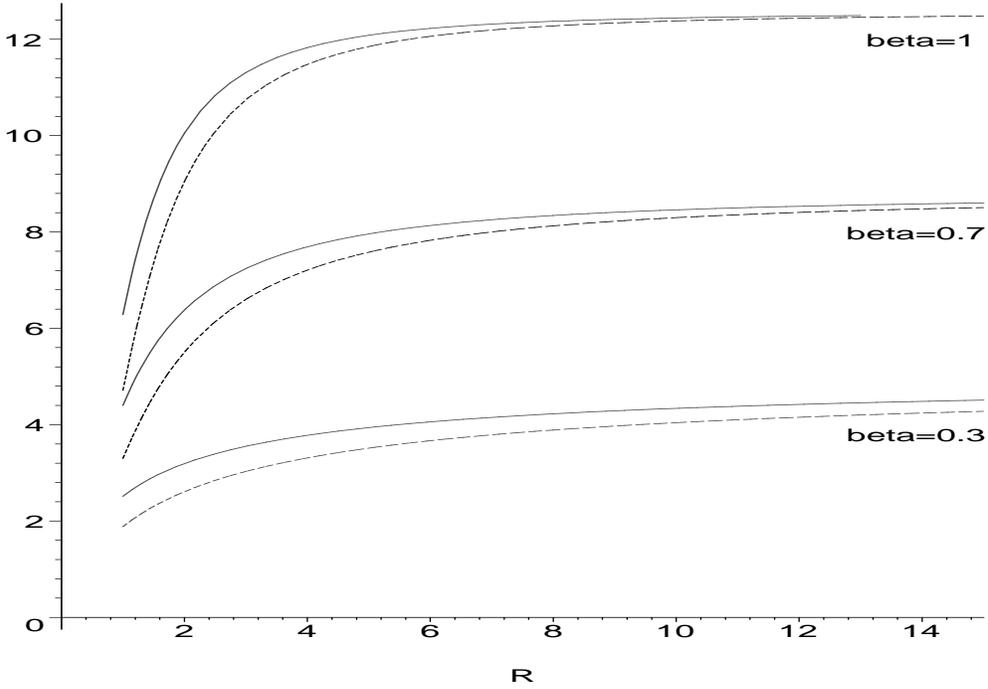}
\vskip -0.0cm \caption{{\protect\small Displays the energies (in units of
$J$-parameter) associated to
$I$ and $P$ solitons, $E_I$ and $E_P$, represented by dashed and solid lines
respectively. Note that as the cone is
tighten, both energies appear to decrease in a such a way that the gap between
$E_I$ and $E_P$ becomes greater, keeping always the more fundamental
characteristic of $I$-type excitations. Here, we have taken $a=1$ and
$c_1=0.1$.}} \label{fig2} 
\end{figure}

How such energies behave as function of soliton radius,
$R$, is displayed in
Figure \ref{fig2}. Similarly to the results presented in
Ref.\cite{PereiraPires}, $I$-type appear to be more fundamental than
$P$-solitons. Indeed, for $\beta=1$, our results exactly recover their ones.
Furthermore, it is worthy noticing that as $\beta$ gets smaller the
associated energies to both kinds of excitations also decrease in such a way
that the gap between $E_I$ and $E_P$ turns out to be larger (see Figure 
\ref{fig2} ). Since the
observation of these solitonic excitations were recently
reported \cite{Subetal}, we claim that similar experiments concerning spins
textures on conical support could determine the energy gap increasing predicted
here.\\

\begin{figure}[ht] 
\centering \hskip 0cm
\includegraphics[width=4.7cm,height=6cm]{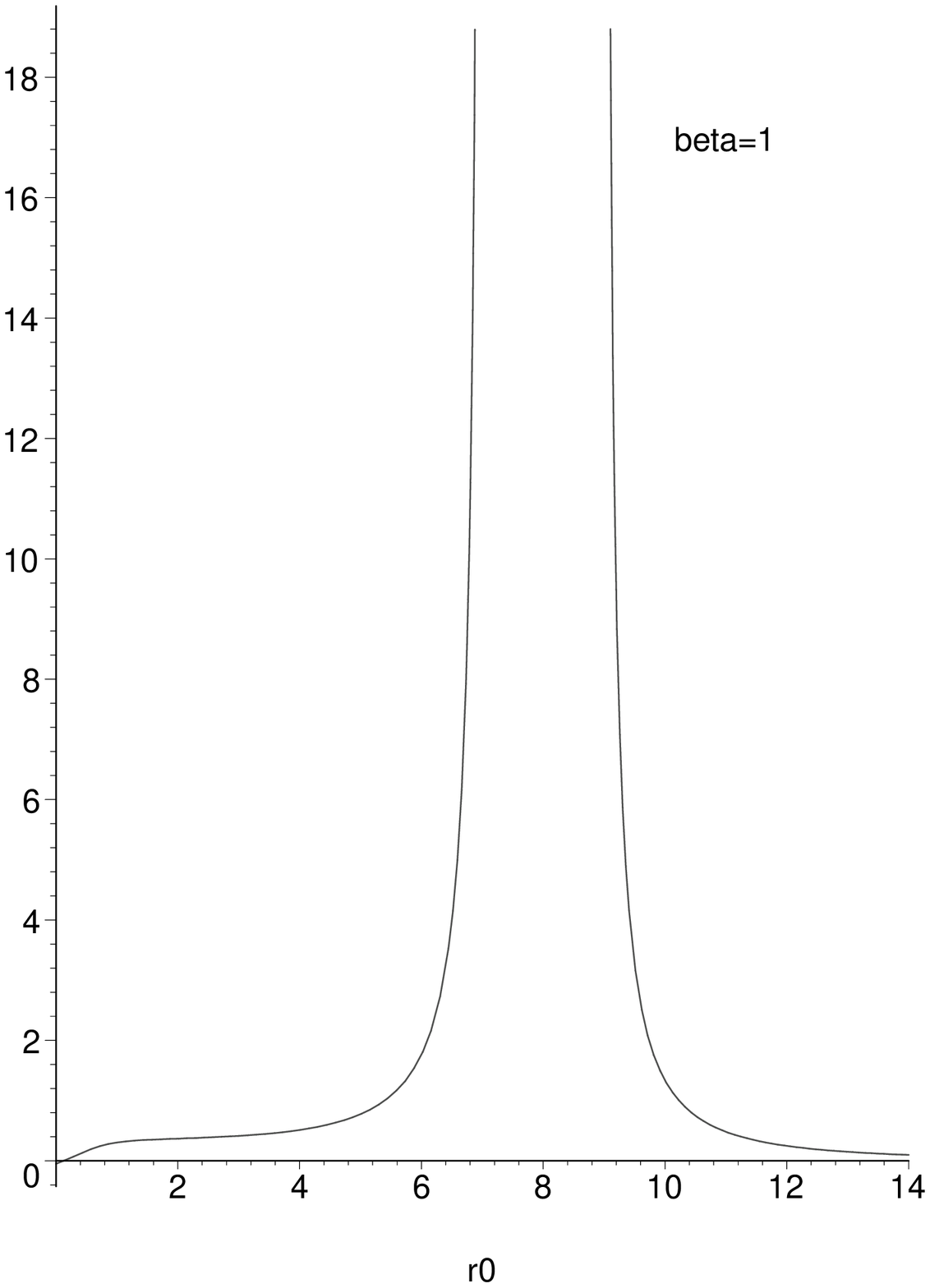}
\hspace{0.6cm}
\includegraphics[width=4.7cm,height=6cm]{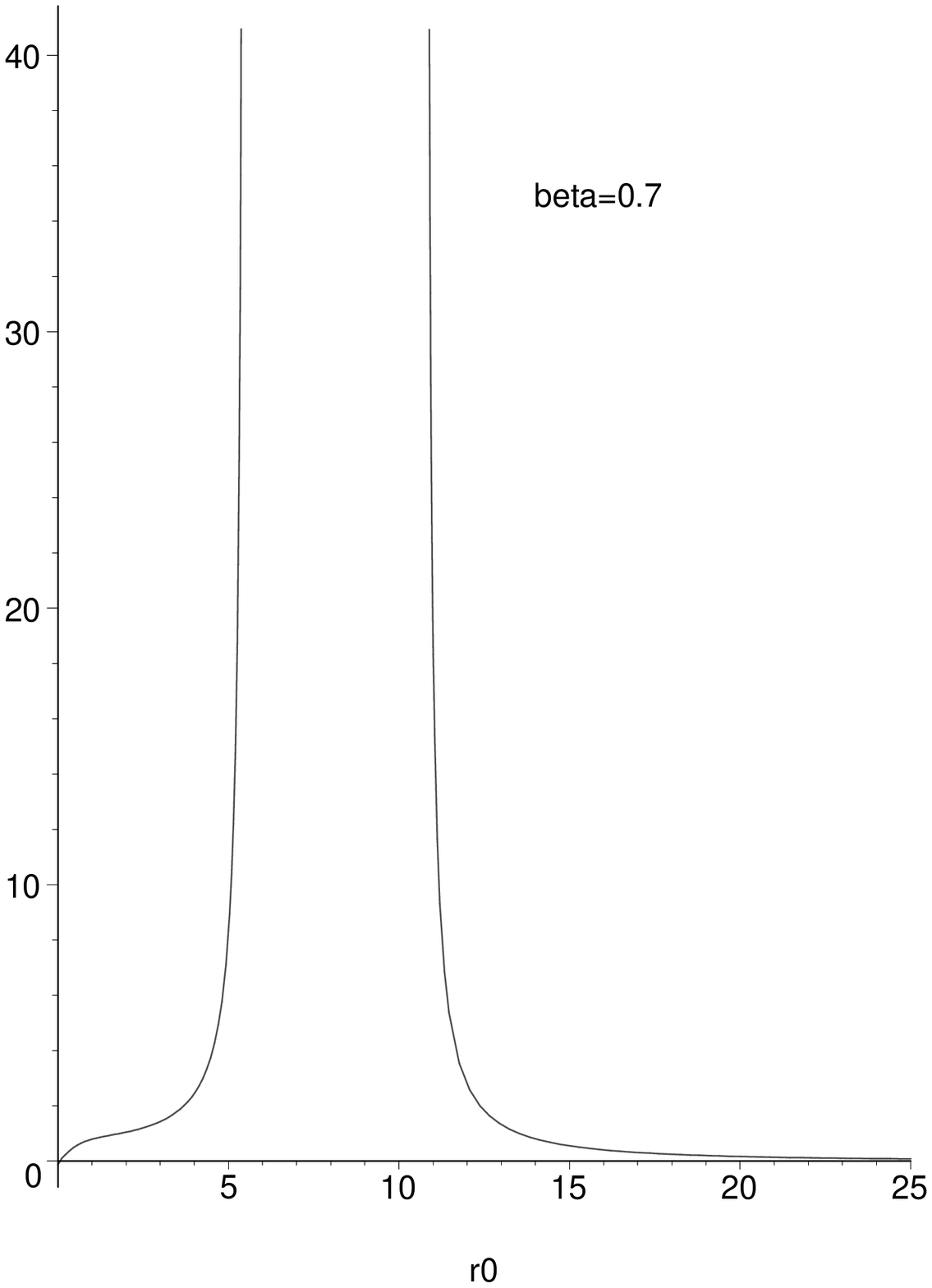}
\hspace{0.6cm}
\includegraphics[width=4.7cm,height=6cm]{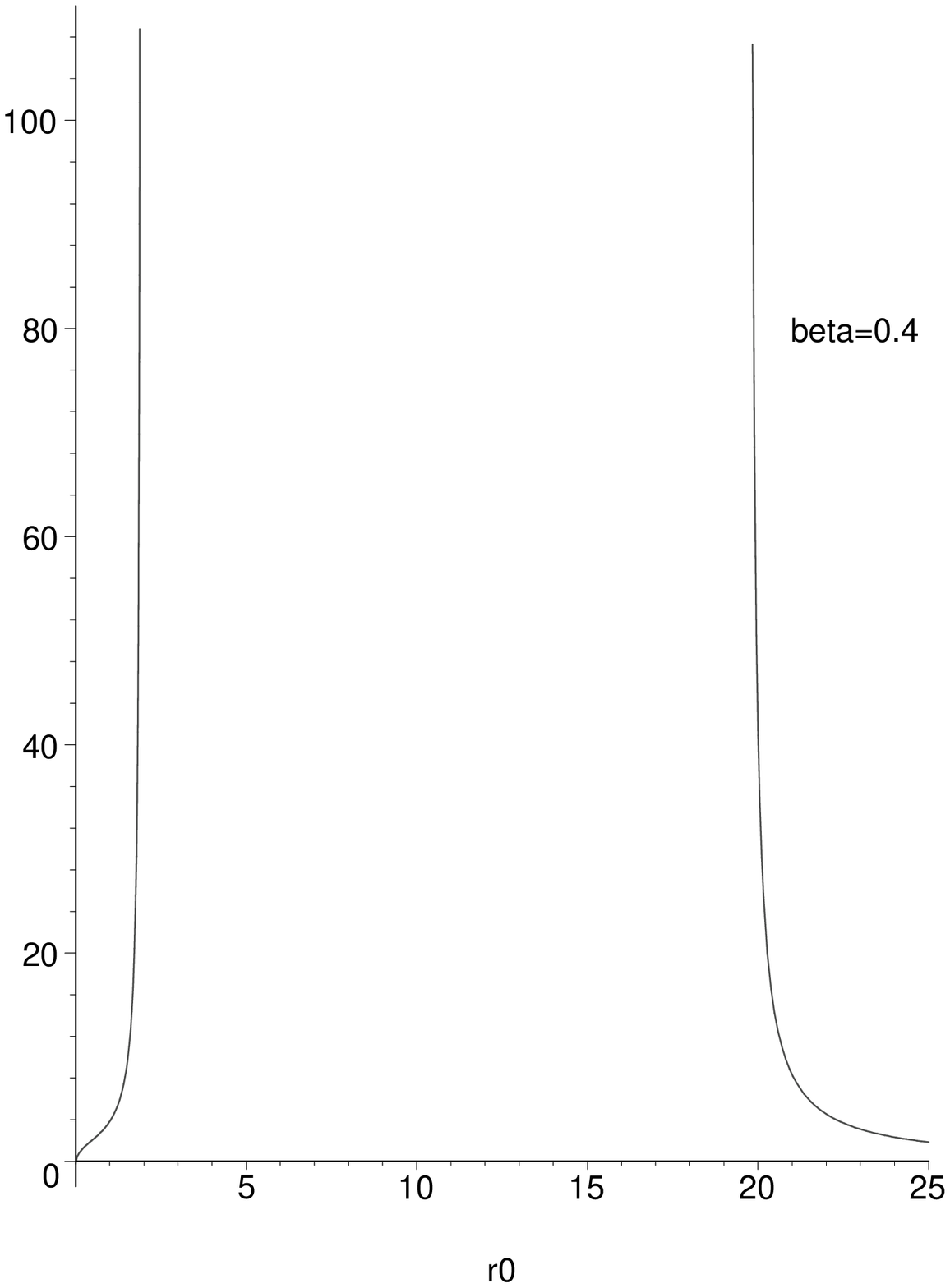}
\vskip -0.0cm \caption{{\protect\small Exhibits the behavior of $U_{\rm eff}$
(in units of $J$-parameter)
as function of the separation soliton-vacancy, $r_0$ (above $R=8a$ and $a=1$).
Note that for short distances the potential is attractive while for large
separations it turns out to be repulsive. Notice also that such an attractive
interval decreases as $\beta$ gets lower values, indicating a stronger
confinement (pinning) of the soliton whenever $\beta$ is sufficiently small.
From left to right, we have $\beta=1,\, 0.7, \,{\rm and}\, 0.4$,
respectively.}} \label{fig3} 
\end{figure}

Actually, in the preceding analysis we have considered that the nonmagnetic
impurity is located at the soliton center, say, $\vec{\rho}=(0,0)$ (the
apex of the cone). However, following the work presented in
Ref.\cite{PereiraPires}, we may wonder whether the effective potential between
them would behave as long as they  were placed one apart other. The way
to
perform that is analogous to that presented in Ref.\cite{PereiraPires}, and we
shall not repeat them here (we refer reader to such a reference for further
details). After some calculation, we get (the soliton is centered at the
apex, $\vec{\rho}=(0,0)=\vec{r}$):

\ba
 & & \hskip -0.7cm U_{\rm eff}=E_0\,\left\{\left[ \frac{R^{2\beta} a^\beta
(r_0^\beta
-a^\beta)^2}{\sqrt{(r_0^\beta -a^\beta)^2 +r_0^{2\beta}}[(r_0^\beta
-a^\beta)^4 -R^{4\beta}]} - \frac{R^{2\beta} a^\beta (r_0^\beta
+a^\beta)^2}{\sqrt{(r_0^\beta +a^\beta)^2 +r_0^{2\beta}}[(r_0^\beta
+a^\beta)^4 -R^{4\beta}]}\right]+ \right.\nn\\
& & \hskip 10cm \left. -\frac14
\left(\frac{(Ra)^\beta}{(r_0^{2\beta}+R^{2\beta})}\right)^2\right\}
\,,\label{Ueff}
\ea

\noindent where $\rho_I=r_0^\beta/\beta$ labels the impurity position. Notice
that the singular points of $U_{\rm eff}$ appear at
$(r_0)^{\mp}_S=(R^\beta\mp\,a^\beta)^{1/\beta}$. For example, if $\beta=1$ and 
$R=5a$, we then have $(r_0)^{-}_S=4a$ and $ (r_0)^{+}_S=6a$. In Figure
(\ref{fig3}) we plot
$U_{\rm eff}$ versus $r_0$ for some values of $\beta$. In general, the
potential is attractive for $r_0<(R^\beta-\,a^\beta)^{1/\beta}$ while for
separation $r_0>(R^\beta+\,a^\beta)^{1/\beta}$ its appears to be repulsive.
Such a behavior was already obtained in the work of
Ref.\cite{PereiraPires}, for the planar, say, $\beta=1$ case.\\

What our results bring as novelty is the fact
that as long as the cone is tighten ($\beta$ gets lower) then the range of the
attractive potential gets shorter. As a consequence, if we fix $R=5a$, and
extrapolate our results to the discrete lattice (with spacing $a$), we
conclude that for $\beta<\beta_{\rm cr}$ ($\beta_{\rm
cr}=\ln(2)/\ln(5)\approx 0.43$, in this case), we get $r_0<a$ what
implies that or the soliton is pinned on the impurity or they are quite apart
one from another thanks to the repulsive potential. On the other hand,
keeping the $R$-value above, whenever $\beta_{\rm cr}<\beta\le1$, we have the
possibility of oscillating solitons around the spin-vacancy. Actually, let us
take $U_{\rm eff}$ above, and consider small displacements impurity-soliton
($r_0$). Expanding it like below:

\ba
U_{\rm eff}|_{r_0\to0}= E_j+\frac12 r_0^2\left(\frac{d^2U_{\rm eff}}{d
r_0^2}\right)_{r_0=0}+\frac13 r_0^3\left(\frac{d^3U_{\rm eff}}{d
r_0^3}\right)_{r_0=0}+\dots\,,
\ea
where we have taken into account the type of the excitation, according to
(\ref{energysolitonI}-\ref{energysolitonP}). Now, looking only for the
harmonic
potential (say, the quadractic term) it is easy to show (we refer the reader
to Ref.\cite{nosprb} for details) that the solitons oscillate around the spin
vacancy with frequencies given by:

\ba
\omega_j=\sqrt{2}\frac{c}{a}\left[ \frac{a_j^{2\beta}}{R^{2\beta}}
+ (R^{2\beta} +a_j^{2\beta})\left( \frac{a^{2\beta}}{\sqrt{5}
[R^{4\beta}- (2a)^{4\beta}]} - \frac{a^{2\beta}}{4(R^{2\beta}
+a^{2\beta})^2}\,, \label{omega} \right) \right]^{1/2}
\ea

\noindent where $a_j=a_I, a_P$ ($a_I=1.01a$ and $a_P=0.23a$, see
Ref.\cite{Subetal}, for
details) while $c=2JSa/\hbar=\sqrt{E_j/M_j}$, where $M_j$ is the soliton
rest mass (for details, see Ref.\cite{Wysin1}).\\

\begin{figure}[ht] 
\centering \hskip -2cm
\includegraphics[width=13cm,height=9cm]{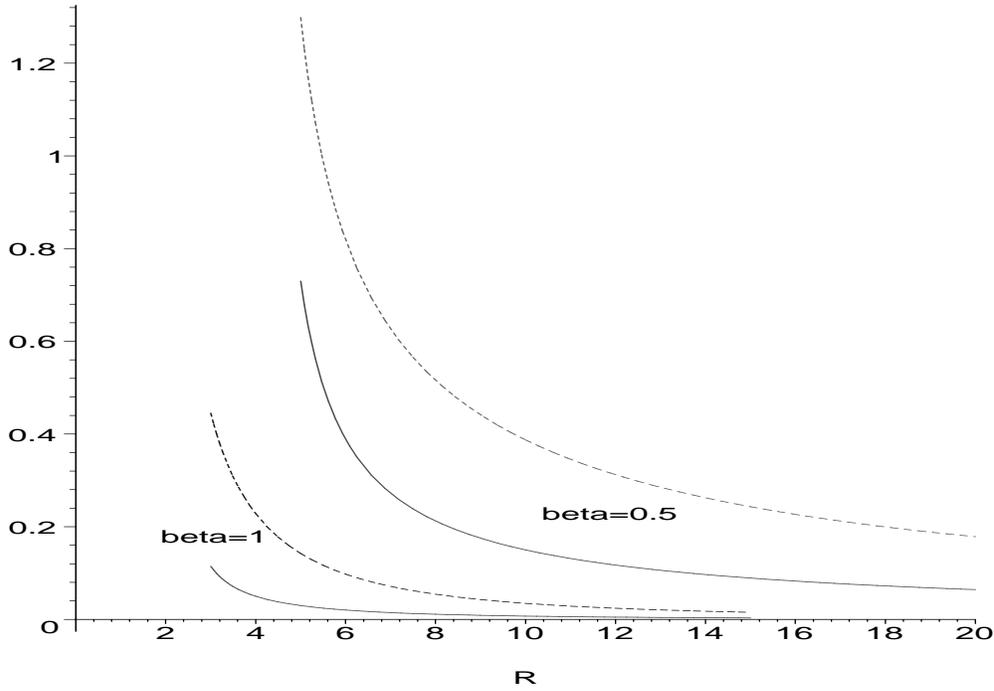}
\vskip -0.0cm \caption{{\protect\small Shows solitonic frequencies, $\omega_j$
(in units of $c/a$),
around the static impurity as function of soliton radius, $R$ (here $a=1$).
Notice that $I$-excitations (dashed lines) present higher frequencies than $P$ ones
(solid lines). Furthermore, note that as long as the cone is tighten, these
excitations oscillate faster (the two lower curves are concerned to $\beta=1$
while the upper ones are related to $\beta=0.5$ ).}} \label{fig4} 
\end{figure}

How $\omega_j$ behaves as function of the type ($I$ or $P$) and soliton size,
$R$, for some values of $\beta$-parameter is shown in Figure (\ref{fig4}).
As expected, and earlier reported in Ref.\cite{nosprb}, as soliton
becomes larger, their frequency modes decrease. In addition, $I$-solitons
oscillate faster than $P$-type ones. The novelty brought about here is that,
as conical support is tighten the excitations appear to oscillate
faster. In addition, notice also that for a given $\beta$ the soliton size
should be $R>2^{1/\beta}a$ (in the discrete lattice). Thus, we see that
for $\beta=1$ we should have $R\ge3a$ while for $\beta=1/2$ such a radius
should take values $\ge5a$. However, despite such a size increasing,
the geometry of the support deeply affects frequencies, as displayed in
Figure (\ref{fig4}).\\

\section{The Zeeman coupling and the instability of static solitonic
solutions}

In this section, we shall briefly discuss how the coupling of spins to an
external magnetic field yields unstable static spin textures. Introducing the
term $(\mu g\vec{B}\cdot\vec{n})\,V(\vec{\rho})=\mu\,gB_0 \,m\,V(\vec{\rho})$
into Hamiltonian (\ref{Hini}), where $\vec{B}=B_0\hat{z}$ ($B_0$ is
homogeneous), the counterparts of dynamical equations
(\ref{eqdphi},\ref{eqdm}) read
like follows:

\ba
& & \hskip -0.8cm \frac{\partial\,m}{\partial\,t}
=V_I(\vec{\rho})\left[\nabla^2\Phi 
-\frac{2m}{(1-m^2)}(\nabla 
m)\cdot(\nabla\Phi)\right]-\nabla\Phi\cdot\nabla 
V_I(\vec{\rho})\,,\label{eqdphiB} \\ & & 
\hskip -1.4cm\frac{\partial\,\Phi}{\partial\,t}= V_I(\vec{\rho})\left[\nabla^2
m +\frac{m}{(1-m^2)}(\nabla 
m)^2 +m(1-m^2)(\nabla\Phi)^2 -\xi_B^{-2}(1-m^2) \right] -\nabla
m\cdot\nabla V_I\,,
\label{eqdmB}
\ea
where $\xi_B^{2}=J/g\mu\,B_0$ is the magnetic length.\\

As we have done in the preceding case, supposing that $J>>g\mu\,B_0$
(spin-spin coupling is much stronger than Zeeman interaction) then, we have
that eq. (\ref{eqfi1}) is still valid here, while (\ref{eqm1}) is modified by
the magnetic field. With the same assumptions taken in the preceding analysis
(impurity at the soliton center, cylindrically symmetric solutions, etc), we
conclude that $\Phi_1={\rm constant}$
remains the solution for its respective differential equation also here.
However, eq. (\ref{eqm1u}) now gets the form (using
$m_1=m_1(m_0)=m_1(u)$ analogously to eq. (\ref{eqm1u})):

\ba
\frac{d^2\,m_1}{du^2}+\frac{2\,m_1}{(1-u^2)}=-\left(\frac{R}{\beta\xi_B}
\right)^{2} \frac{(1-u^2)^{\frac{1-\beta}{\beta}}}{(1-u)^{2/\beta}} 
\label{eqm1uB}
\,.
\ea

\begin{figure}[ht] 
\centering \hskip -2cm
\includegraphics[width=14cm,height=9cm]{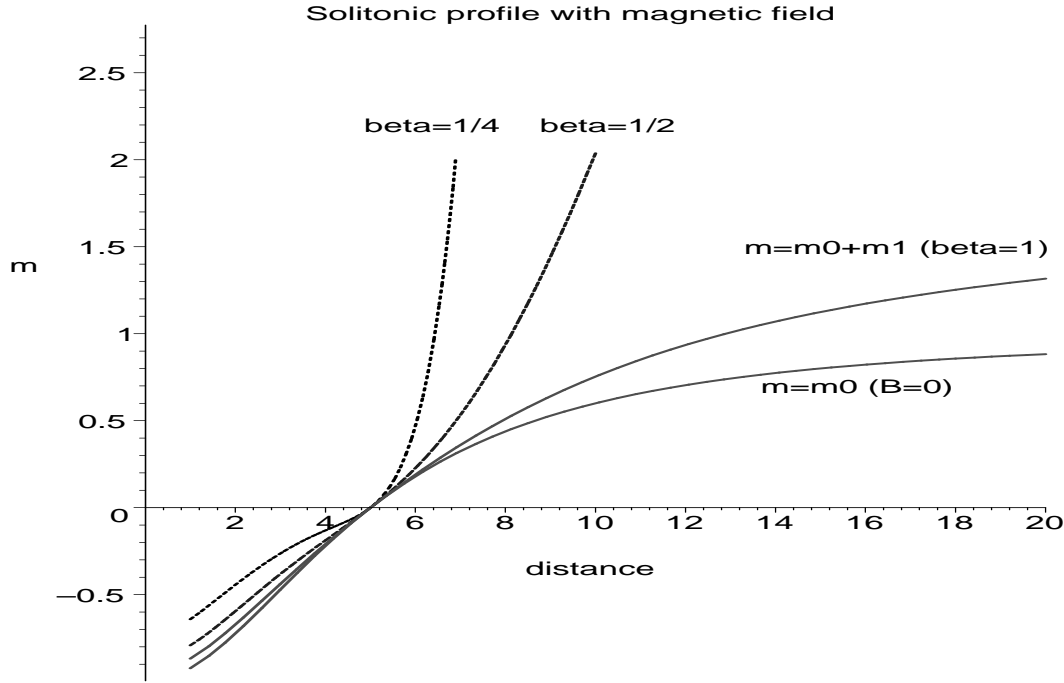}
\vskip -0.0cm \caption{{\protect\small Displays $m$ solitonic profile versus
distance, $r$. Without external magnetic field, $m=m_0$ and $m\to1$ as
$r\to\infty$. However, whenever $\vec{B}$ is turned on, such a profile
present values $>1$ for $r>L_c$, what contradicts $m$ definition and implies
that static solitons becomes unstable. In addition, note that such a critical
size, $L_c$ gets smaller values as $\beta$ is lowered.}} \label{fig5} 
\end{figure}

Contrary to eq. (\ref{eqm1u}), the above one does not present analytical
closed solutions for arbitrary $\beta$, as far as we have tried. Indeed, by
using {\em MAPLE8} we have obatined closed expressions for some
$\beta=1/N$-values ($N$ positive integer). Namely, we have noted that such
expressions becomes more complicated as $N$ is rised. By virtue of their
length, we shall not write them here. Rather, in Figure
(\ref{fig5}) we plot $m=m_0+m_1$ as function of distance, $r$. The main
point to be noticed here is that no good $m_1$-functions solves eq.
(\ref{eqm1uB}), what may be realized by noting that $m$ blows up as
$r\to\infty$, yielding divergent associated energy. Actually, in the
very beginning we have defined $m=\cos(\theta)$, but the results plotted
in Figure
(\ref{fig5}) explicitly shows us that for $r>L_c$ (depending upon $\beta$)
$m$-variable takes values $>1$, what is a contradiction. Thus, we conclude
that whenever Zeeman corrections are taken on $m$ solitonic profile these
excitations appear to get unstable and collapse. Furthermore, once that such  a
blowing up character of $m$-profile comes about by turning on
$\vec{B}$-field and appear to become stronger as long as the cone aperture
gets smaller, we may realize that by decreasing $\beta$-parameter we can
increase the net magnetic effect on solitonic excitations. Such a result can 
be viewed as a geometrical accentuation of the magnetic field
experienced by solitons.\\

Notice, however, that we
cannot state that they shrink to point-like objects, since the impurity
introduces a characteristic length and no smaller size than spin vacancy is
possible (recall that the impurity was placed at soliton center). In the
discrete lattice case, we would expect that Zeeman
interaction should force solitons to shrink to the smallest size: the
lattice spacing $a$. Notice, however, that this is pointed out here as a
possibility (following the conclusions presented in
Refs.\cite{curvatura1,curvatura2}) once our analysis is strictly valid only
for $r>a$. Indeed, a similar problem was earlier considered in the work of
Ref.\cite{curvatura2}. There, the authors explicitly state the impossibility
of carrying out their analytical analysis for the highly complicated
diferential equations.\\

Even though we could not provide here a rigorous proof of whether the
solitons necessarily decreases their radius to spin vacancy size in order to
compensate the instability brought about by Zeeman interaction, we should
mention that the present problem is important in connection with dynamical
solitons, for instance, in the lines of
Ref.\cite{dynsolitons}. Thus, considering precessional
oscillatory modes of a soliton around $\vec{B}$, their associated frequencies 
could be determined and how geometrical features affect such modes. Such an
issue is under investigation and results will be communicated
elsewhere\cite{nosworkinprogress}.

\section{Conclusions and Prospects} 

We have considered the problem of classical Heisenberg 
spins on a circular conical geometry subject to both the 
potential of a static non-magnetic impurity and an 
external magnetic field. Whenever spins weekly interact with the
impurity alone, a number of results were obtained and compared to their planar
counterparts. In general, our results recover planar ones as long as
$\beta=1$, and for arbitrary cone aperture they describe how solitonic spin
textures are affected by a static spin vacancy in such a geometry. Among
other, we have seen that the impurity presence now allows the appearance of two
kinds of  solitons: $I$ and $P$-type ones. In this line, one interesting thing
realized in conical geometry is that the energies of both types get smaller
values whenever the cone is tighten while the energy gap between them is
increased. Since both kinds of excitations have been recently observed in
experiments, we could think of similar experiments dealing with Heisenberg-like
spins on conical supports in order to determine the results predicted here.\\

Since NL$\sigma$ model has been also applied to study weekly
self-interacting two-dimensional electron gas (2DEG) \cite{Sondhietal}, our
present study may also be relevant, for instance, in connection with Quantum
Hall Effect in curved, say, conical, geometry.\\

An analysis of how external magnetic field influences the structure of
dynamical solitons is in order. For instance, how their
internal precessional modes and sizes are influenced by this field are
some of the issues we have been addressing. Results concerning such a study
will be communicated elsewhere\cite{nosworkinprogress}.\\

As prospects for future investigation, we may quote the introduction of more
impurities in the system, for example, as presented in the works of
Refs.\cite{Moletalprb2003,Fagneretal}. There, analytical and
numerical/simulational analysis of the model may indicate how collective spin
vacancies effects are read by solitonic excitations.\\

Concerned with the current topic of magnetic nanostructured objects, how
nanomagnetic cones \cite{Ross1,nanocones} (among other curved
structures), with and without vacancies and magnetic field, affects the
magnetization and hysteresis loops are some interesting points to be
studied.\\  

Finally, in connection with Field Theory/High Energy Physics and 
related topics, a problem that could lead to interesting 
results is that of a charged particle interacting with a 
magnetic monopole in a conical background. Perhaps, in this 
framework, bound states between them could be observed, 
scenario quite distinct from its flat space counterpart, 
which does not present closed orbit configurations between the
m \cite{Dirac48}.\\

\vskip 1cm
{\Large\bf \centerline {Acknowledgments}}\vskip .5cm
WAF and ARP  thank CNPq for financial support.  WAM-M acknowlegdes
FAPEMIG for partial financial support.

\end{document}